# Effect of Hydrostatic Pressure on $BiS_2$-Based Layered Superconductors: A Review

Rajveer Jha and V.P.S. Awana[*]

CSIR-National Physical Laboratory, Dr. K.S. Krishnan Marg, New Delhi-110012, India

**Abstract:**

We report impact of hydrostatic pressure on the newly discovered $BiS_2$ based superconductors. In last couple of years the new $BiS_2$ based superconductor attracted great attention of the condensed matter physics community. These new layered $BiS_2$-based compounds are very sensitive to the concentration of carriers doping and pressure that cause profound changes in their physical properties with appearance of superconductivity in the vicinity of their insulating/semiconducting state. The $BiS_2$-based new compounds are expected to provide us with the next stage of exploring new superconductors and to discuss their exotic superconductivity mechanisms. In current review, we present most of our findings related to impact of hydrostatic pressure on superconductivity of new $BiS_2$ based superconductors at one place.

[*]**Corresponding Author**
Dr. V. P. S. Awana, Principal Scientist
E-mail: awana@mail.nplindia.org
Ph. +91-11-45609357, Fax-+91-11-45609310
Homepage www.fteewebs.com/awanavps/

## Contents

# 1. Introduction

Superconductivity was first found in the $Bi_4O_4S_3$ compound with the $T_c$ value of 4.5K in the year 2012 by Y. Mizuguchi et.al [1]. Within a week or so, S. K. Singh et al. reproduced the superconductivity and confirmed the bulk characteristics of the emerging superconductivity from transport and magnetic measurements [2]. Structurally, thecompound is the layered one [1-6]. In the same year, other $BiS_2$ based superconductors have been discovered like $LaO_{1-x}F_xBiS_2$, $CeO_{1-x}F_xBiS_2$, $PrO_{1-x}F_xBiS_2$ and $NdO_{1-x}F_xBiS_2$ with $T_c$ of 3.0, 2.5, 3.5 and 5.6K respectively [7-17]. These compounds are similar to the Cuprates [18] and Fe-based superconductors [19] in layered crystal structure and some exotic physical properties. As a matter of fact, the parent phase of these compounds, i.e., $CeOBiS_2$ has been synthesized earlier in 1976 itself [20]. Later in 1997, some other similar compounds viz. $REOBiS_2$ (RE=La, Pr, Nd and Yb) were synthesized, and found to exhibit semiconducting behavior [21]. Superconductivity in these compounds was observed by electron doping via $F^{1-}$ substitution at the s $O^{2-}$ site [7-17]. The superconductivity in $BiS_2$ based compounds has also been observed via substitution of tetravalent $Th^{+4}$, $Hf^{+4}$, $Zr^{+4}$, and $Ti^{+4}$ for trivalent $La^{+3}$ in $LaOBiS_2$ [22]. Superconductivity was also reported in $SrFBiS_2$ compound, via electron doping by substitution of $La^{3+}$ at the site of $Sr^{2+}$ [23,24]. The $SrFBiS_2$ seems to be the parent compound of the 1112 family [24]. The newest layered $BiS_2$-based superconducting systems appear to be very sensitive to the carrier doping level, as the atomic substitutions cause profound changes in their properties [25-33]. The $BiS_2$-based compounds are expected to provide us with the next stage to explore new superconductors and discuss the exotic superconductivity mechanisms. The electronic structure of $Bi_4O_4S_3$ has been discussed from theoretical band calculations for both the parent and superconducting phases [1]. The electronic structure of $BiS_2$ based superconducting materials has been proposed by several groups by using theoretical calculations like density functional theory [38-56]. X. Wan et. al., compared the experimental and theoretical results, and found that changing the interlayer distances only slightly affects the band structure [41]. For $LaOBiS_2$, both S$3p$ and O$2p$ states appear mainly between −4.0 and 0.0 eV. Although it is located primarily above the Fermi level, Bi $6p$ also makes a considerable contribution to the states between −4.0 and 0.0 eV, indicating a strong hybridization between Bi-$6p$ and S-$3p$ states [56]. The crystal structure of $LaO/FBiS_2$ is composed of a stacking of LaO/F layers and double $BiS_2$ layers. In terms of electronic bands, the parent compound $LaOBiS_2$ is consisting of valence band O and S $p$ states and a conduction band of Bi-*6p* and S-*3p* states, which is a band insulator with an energy gap of 0.4 eV [38,41].S. Nagira et al. obtained core level spectrum and showed the appearance of a new spectral component on the lower binding energy side of Bi 4f, which may be explained by core hole screening with metallic states near $E_F$. The electronic structure information on the new superconductor by Valence band SXPES measurements including the shape of the valence band, x-dependent energy shift, and appearance of states at $E_F$ with the Bi 6p character, which show overall agreement with band structure calculations. [57]. L. K. Zeng et al. ARPES results show the anomalous temperature dependence of the low-energy spectrum indicates that the electrons could be strongly coupled with the lattice in the low-temperature normal state of this superconductor. [58]. While Z. R. Ye et al. and X. B. Wang et al. found small Fermi pocket size and the weak electron correlations and suggest that the $BiS_2$-based superconductors could be conventional BCS superconductors mediated by electron phonon coupling [59,60].T. Machida et al. report the origin of the superconductivity in $NdO_{0.7}F_{0.3}BiS_2$ system by STM and STS experiments, and say that the electronic spatial structures coexist with the bulk superconductivity [61]. S. F. Wu et al. performed polarized Raman scattering measurements on the newly discovered superconductor $Nd(O,F)BiS_2$ and suggest that the $BiS_2$-based superconductors are possibly phonon-mediated BCS superconductors [62]. H.-F. Zhai et al. report another phase $EuFBiS_2$ and $Eu_3Bi_2S_4F_4$ of $BiS_2$ based superconducting compounds [63, 64].

The role of rare earth elements is also crucial in superconductivity of $BiS_2$ based compounds as in the Fe based and High $T_c$ cuprate superconductors. Some experimental studies show that superconductivity effectively increases as light rare earth element replaced by heavier rare earth element [7-18]. Interestingly, chemical substitutions play an important role to induce superconductivity in the $BiS_2$ based compounds. The external pressure dependent studies have also been done effectively on these compounds, which modified the superconductivity of $BiS_2$ based superconductors [65-80]. Due to the applied pressure the change in lattice parameters through bond lengths and angles took place, which positively affected the electronic and magnetic correlations of the superconductor [81].In addition, the structural phase transition from tetragonal to monoclinic has been explained on the basis of analysis of the x-ray diffraction patterns under high pressure of the $LaO_{0.5}F_{0.5}BiS_2$ compound and perfect tetragonal phase has been reported for the high pressure synthesized of $CeO_{0.3}F_{0.7}BiS_2$ compound [69,70].

## 2. Effect of hydrostatic pressure
### 2.1 $REO_{1-x}F_xBiS_2$ (RE=La, Ce, Pr and Nd)

The first pressure study on for $Bi_4O_4S_3$ compound was investigated by Kotegawa et al. up to applied pressures 1.92 GPa [65]. The $T_c$ gradually decreases without a distinct change in the metallic behavior in the normal state resistivity [65]. The same has been reported by Kalai et al. [66]. Superconductivity above the 10K for high pressure synthesized $LaO_{0.5}F_{0.5}BiS_2$ has been reported under hydrostatic pressure by Kotegawa et al. [65] and further the evolution of $T_c$ under applied pressure for $LnO_{0.5}F_{0.5}BiS_2$ samples were reported by Wolowiec et al. [66–71].

Now after well establishing the superconducting properties $REO_{0.5}F_{0.5}BiS_2$ (RE=La, Ce, Pr and Nd) superconductors [7-17,100], we study the impact of hydrostatic pressure on their superconductivity. In our previous studies the temperature dependent electrical resistivity from 300K down to 2K at various applied pressures for the $REO_{0.5}F_{0.5}BiS_2$ compound see figure 1. At ambient pressure normal state resistivity $\rho(T)$ exhibits semiconducting behavior for all the compounds, which is strongly suppressed with increasing pressure from 0-0.97GPa and then suppressed systematically before being practically unchanged for the higher applied pressures of above 1GPa for $LaO_{0.5}F_{0.5}BiS_2$ and $CeO_{0.5}F_{0.5}BiS_2$ compounds see figure 1(a, b). While the normal state electrical resistivity increases for higher applied pressure above the 0.55GPa for the $PrO_{0.5}F_{0.5}BiS_2$ and $NdO_{0.5}F_{0.5}BiS_2$ compounds see figure 1(c, d). Figure 2(a-d) is the clearly visualize the superconducting transition temperature ($T_c$) of $REO_{0.5}F_{0.5}BiS_2$ compounds for the various applied pressures. The superconducting temperature $T_c$ for $LaO_{0.5}F_{0.5}BiS_2$ can be seen from the Figure 2a $T_c(\rho=0)$ is 2K at 0GPa pressure, which slightly increases to 2.5K at applied pressure of 0.35GPa. Interestingly for 0.55GPa applied pressure, the superconducting transition temperature $T_c(\rho=0)$ increases to 5K with $T_c^{onset}$ at 9.8K and the transition broadening $[T_c^{onset}-T_c(\rho=0)]$ is significantly high. For the applied pressure of 0.97GPa the $T_c(\rho=0)$ is increased to 9.8K with $T_c^{onset}$ at around 10.5K, having a relatively sharp transition. The estimated pressure coefficient $dT_c(onset)/dP$ of $LaO_{0.5}F_{0.5}BiS_2$ is +8K/GPa. Increase in the $T_c$ with applied pressure for $CeO_{0.5}F_{0.5}BiS_2$ is shown in figure 2b. It is clear from figure 2b, at zero pressure $T_c(\rho=0)$ is 2.2K and which is unchanged for 0.35GPa pressure, but the same increases slightly to 2.5K at 0.55GPa pressure. At 0.97GPa pressure the $T_c(\rho=0)$ increases to above 3.5K with a slightly broader transition width. At further higher pressure of 1.38GPa the $T_c(\rho=0)$ is increased to 6K.

For still higher applied pressures of 1.68 and 1.97GPa the $T_c(\rho=0)$ is seen at 7K with 8K onset $T_c$. The pressure coefficient $dT_c(onset)/dP$ of $CeO_{0.5}F_{0.5}BiS_2$ is estimated to be near 6K/GPa.

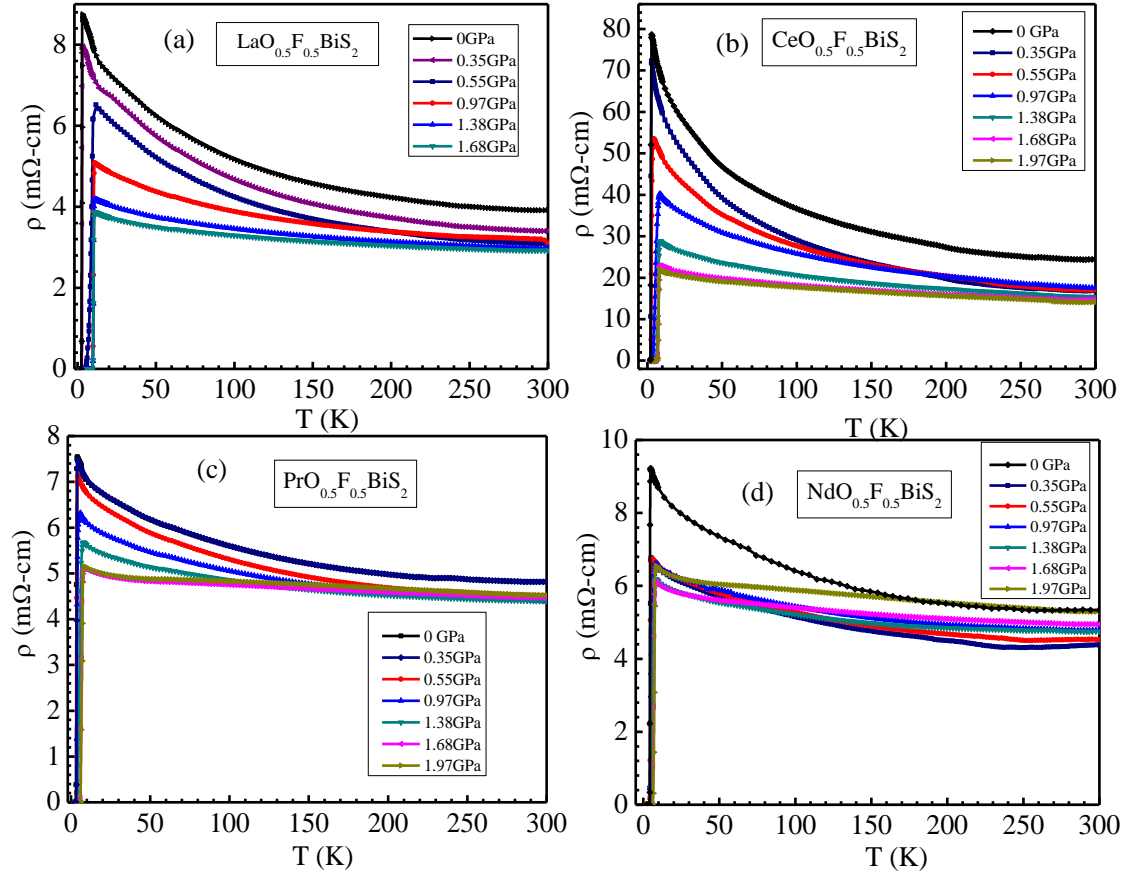

**Figure 1:** (a-d) ρVs T plots for $REO_{0.5}F_{0.5}BiS_2$ (RE=La, Ce, Pr and Nd) compounds, at varying pressures in the temperature range 300K-2K.

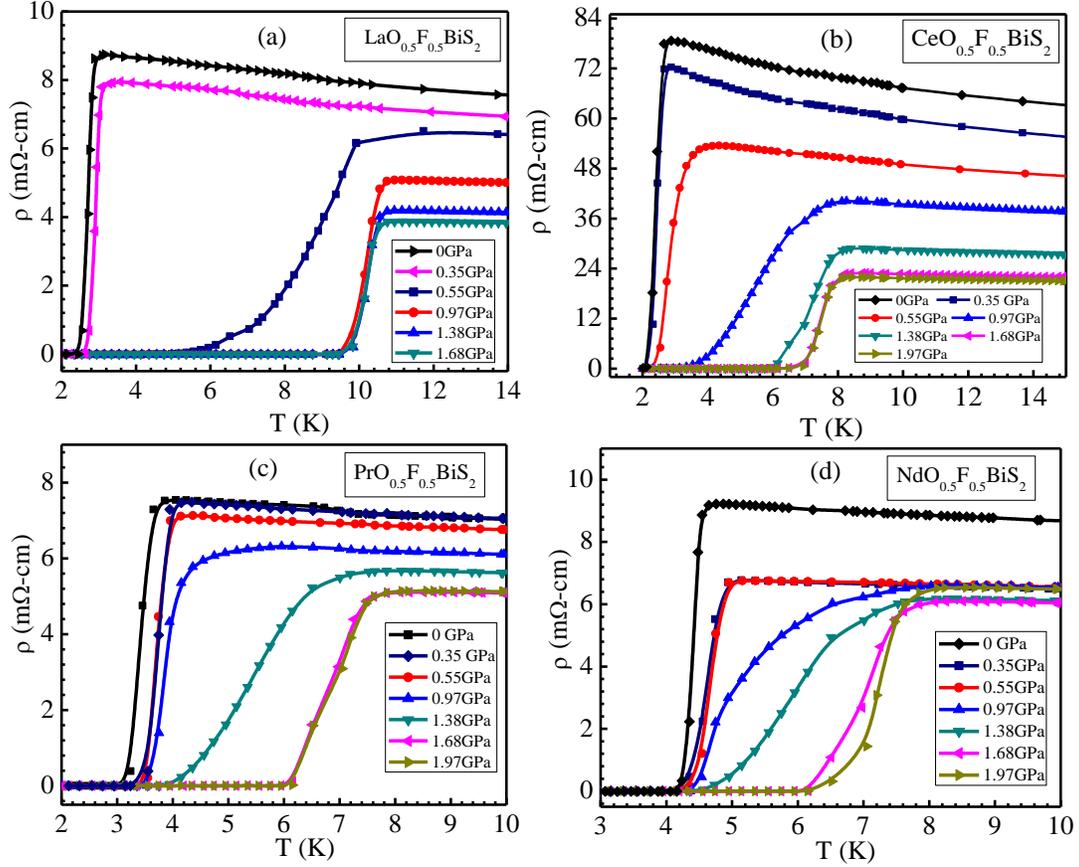

**Figure 2:** (a-d) ρ Vs T plots for REO$_{0.5}$F$_{0.5}$BiS$_2$ (RE=La, Ce, Pr and Nd) compounds, at various applied pressures in the T$_c$ range.

It is clearly seen from the figure 2c with applied pressures the superconductivity increases from 3.5K to up to 7.8K. The superconducting temperature T$_c$ slightly increases 3.5K to 4K for the applied pressure from 0 to 0.97GPa. For 1.38GPa pressure the T$_c^{onset}$ increases to above 6K with the broadening of transition. For further higher pressures of 1.68 and 1.97GPa, the T$_c^{onset}$ increases up to 7.8K with sharp superconducting transitions. The estimated dT$_c^{onset}$/dP of PrO$_{0.5}$F$_{0.5}$BiS$_2$ is around 2.5K/GPa for the obtained highest T$_c^{onset}$ of 7.8K for the applied pressure 1.68GPa. Figure 2d shows the T$_c$=4.3K at 0GPa pressure and slightly increases up to T$_c$=4.8K for the applied pressure 0.35-0.55GPa. At the applied pressure 0.97GPa the superconducting transition temperature T$_c$(ρ=0) increases to 5K with T$_c^{onset}$=6K and the transition broadening is significantly high. For the applied pressure of 1.38GPa the superconducting transition temperature is nearly the same but the normal state resistivity behavior changed. For the further higher applied pressure 1.38GPa the T$_c$ of the sample is changed as T$_c^{onset}$=7.5K, for the 1.68GPa pressure the T$_c^{onset}$ is unchanged at pressure 1.97GPa but the resistivity at low temperature is increases at this pressure. The pressure coefficient $dT_c(onset)/dP = 1.90 K/GPa$ of $NdO_{0.5}F_{0.5}BiS_2$ is comparably lower from the above studied both compounds.

The Pressure dependent superconducting transition temperature $T_c^{onset}$ and $T_c(\rho=0)$ for the $LaO_{0.5}F_{0.5}BiS_2$, $CeO_{0.5}F_{0.5}BiS_2$, $PrO_{0.5}F_{0.5}BiS_2$ and $NdO_{0.5}F_{0.5}BiS_2$ compounds has been shown in the Figure 3(a, b). For the $LaO_{0.5}F_{0.5}BiS_2$ compound the $T_c$ onset is enhanced sharply at the pressure 0.55GPa, while $T_c(\rho=0)$ increases up to 5K. It may be due to the mixed state of lower and high $T_c$ phase of the $LaO_{0.5}F_{0.5}BiS_2$ compound. At the applied pressure 0.97GPa both the $T_c^{onset}$ and $T_c(\rho=0)$ shifts to the 10K and 9.8K respectively, which is the transition pressure ($P_t$) of $LaO_{0.5}F_{0.5}BiS_2$ compound. For the $CeO_{0.5}F_{0.5}BiS_2$ compound at 0.97GPa pressure the $T_c^{onset}$ increases up to the 7K and the $T_c(\rho=0)$ increases to above 3.5K. For the further higher pressure of 1.38GPa though the $T_c^{onset}$ increases to around 8K only and $T_c(\rho=0)$ is increased sharply from 3.5K to 6K. For the $PrO_{0.5}F_{0.5}BiS_2$ and $NdO_{0.5}F_{0.5}BiS_2$ compounds the $P_t$ is at 1.68GPa. For the $LaO_{0.5}F_{0.5}BiS_2$ sample the superconductivity initially increase up to 10K at the 0.97GPa and then it's saturate for the further higher pressures of 1.38GPa and 1.68GPa. In our results we did not observed the decreases in superconductivity up to the applied pressure of 1.68GPa in $LaO_{0.5}F_{0.5}BiS_2$ compound. For the $PrO_{0.5}F_{0.5}BiS_2$ and $NdO_{0.5}F_{0.5}BiS_2$ compounds the superconductivity gradually increases, we observed maximum $T_c(\rho=0)$ of 6.2K at pressure 1.68GPa, which is saturated at the same value up to 1.97GPa pressure.

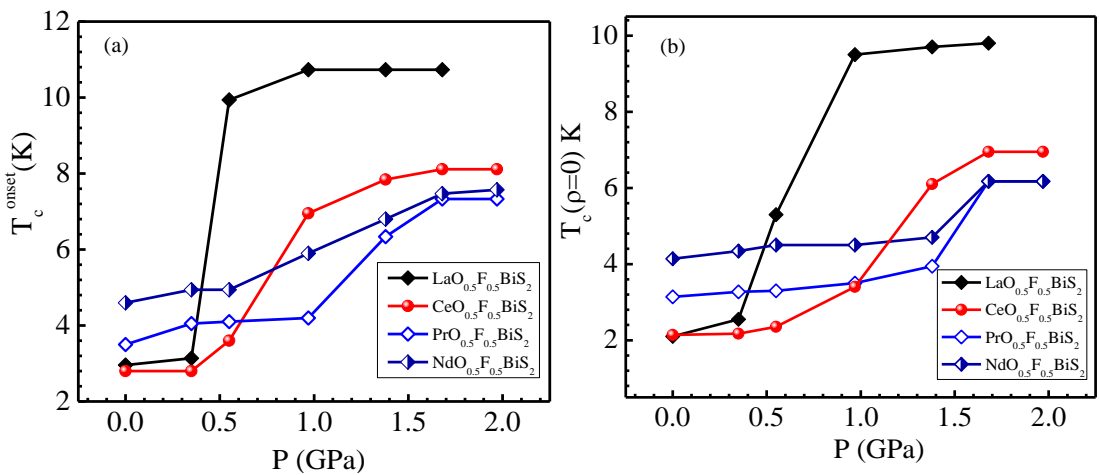

**Figure 3:** (a, b) $T_c^{onset}$ and $T_c(\rho=0)$ versus pressure plots for $LaO_{0.5}F_{0.5}BiS_2$, $CeO_{0.5}F_{0.5}BiS_2$, $PrO_{0.5}F_{0.5}BiS_2$ and $NdO_{0.5}F_{0.5}BiS_2$ compounds.

Figure 4 shows the log(ρ) vs 1/T plots of $LaO_{0.5}F_{0.5}BiS_2$ and $CeO_{0.5}F_{0.5}BiS_2$ compounds at the various applied pressure to find out the values of energy gaps in the two different temperature ranges. It can be seen from the resistivity under pressure data that the normal state electrical resistivity decreases with increasing pressure. It has also been observed from ρ(T) data that the strong suppuration of normal stare resistivity is seen till 0.97GPa and weaker suppuration later for the higher applied pressure for both the compounds. The resistivity data can be described in to distinct regions from the relation $\rho(T)=\rho_0 e^{\Delta/2k_BT}$ where $\rho_0$ is a constant, Δ is an energy gap and $k_B$ is Boltzmann constant. Resistivity data in the whole temperature range is not possible to fit linearly in above relation so we are using the similar trained which has been used by Kotegawa et al. [65]. They explain the two energy gap as $\Delta_1$ and $\Delta_2$ in the temperature range 300 to 200K and from 20K to $T_c$ respectively. The estimated values of the energy gaps are $\Delta_1/k_B \approx 201K$ and $\Delta_2/k_B \approx 11.98K$ for the $LaO_{0.5}F_{0.5}BiS_2$ compound at zero pressure and $\Delta_1/k_B \approx 2024.55K$ and $\Delta_2/k_B \approx 65.68K$ for the $CeO_{0.5}F_{0.5}BiS_2$ compound. The obtained values are comparable to the previous report on the same compounds [67] and significantly less than the values reported by Wolowiec *et al.* for the $LaO_{0.5}F_{0.5}BiS_2$ and $CeO_{0.5}F_{0.5}BiS_2$ compounds [68].

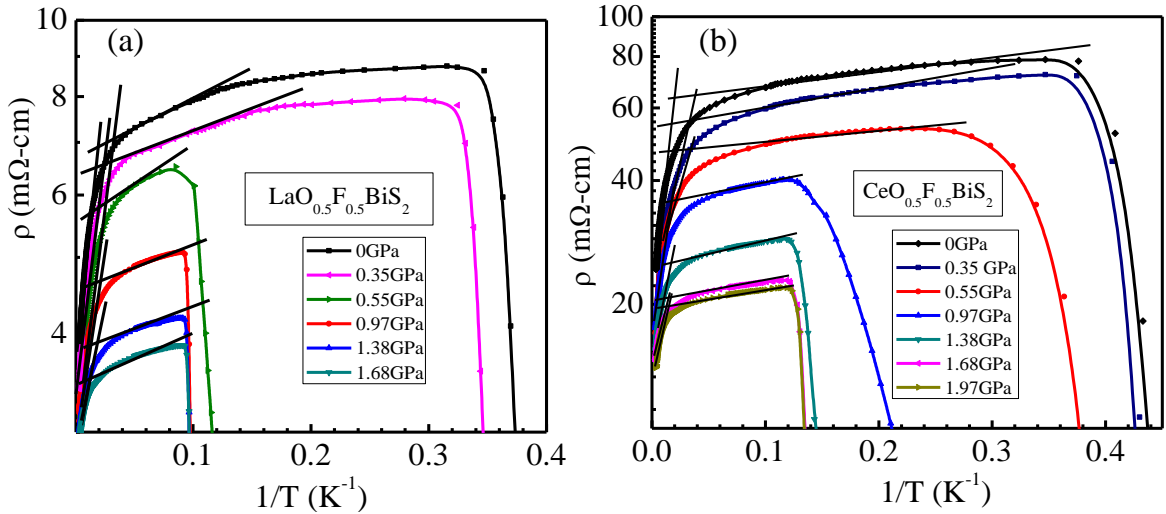

**Figure 4:** (a) ρ vs 1/T plots up to 1.68 GPa for $LaO_{0.5}F_{0.5}BiS_2$ compound and (b) ρ vs 1/T plots up to 1.97 GPa for $CeO_{0.5}F_{0.5}BiS_2$ compound. The solid lines are liner fit of $\rho(T)=\rho_0 e^{\Delta/2k_BT}$ equation at high and low temperatures.

Figure 5(a-d) shows the behavior of energy gaps at the different pressures from 0-1.68GPa for $LaO_{0.5}F_{0.5}BiS_2$ and 0-1.97GPa for $CeO_{0.5}F_{0.5}BiS_2$, $PrO_{0.5}F_{0.5}BiS_2$ and $NdO_{0.5}F_{0.5}BiS_2$

compounds. Both the energy gaps $\Delta_1$ and $\Delta_2$ decreases with the applied pressure from 0-1.68GPa saturation in the decrement of energy gaps may obtained for the higher applied pressure, which has been reported by Wolowiec *et al.* [67]. Figure 5(b) shows the variation of calculated energy gaps ($\Delta_1/k_B$, $\Delta_2/k_B$) with various applied pressures from 0-1.97GPa. It can be clearly seen from Figure. 5(b) that both energy gaps $\Delta_1$ and $\Delta_2$ decrease rapidly with applied pressure of up to 1.68GPa and are almost saturated for 1.97GPa pressure. We have also obtained the values of energy gaps $\Delta_1$ and $\Delta_2$ for the $PrO_{0.5}F_{0.5}BiS_2$ & $NdO_{0.5}F_{0.5}BiS_2$ samples. At the ambient pressure the evaluated values of energy gaps are $\Delta_1/k_B \approx 108.3K$ and $\Delta_2/k_B \approx 4.7K$ for the $PrO_{0.5}F_{0.5}BiS_2$ compound and $\Delta_1/k_B \approx 160.98K$ and $\Delta_2/k_B \approx 6.85K$ for the $NdO_{0.5}F_{0.5}BiS_2$ compound. From the Figure 5 (c) and (d) it can be observed that the both energy gaps decreases rapidly with increasing the pressure. For the $PrO_{0.5}F_{0.5}BiS_2$ compound the first energy gap value at the 1.97GPa is almost saturate.

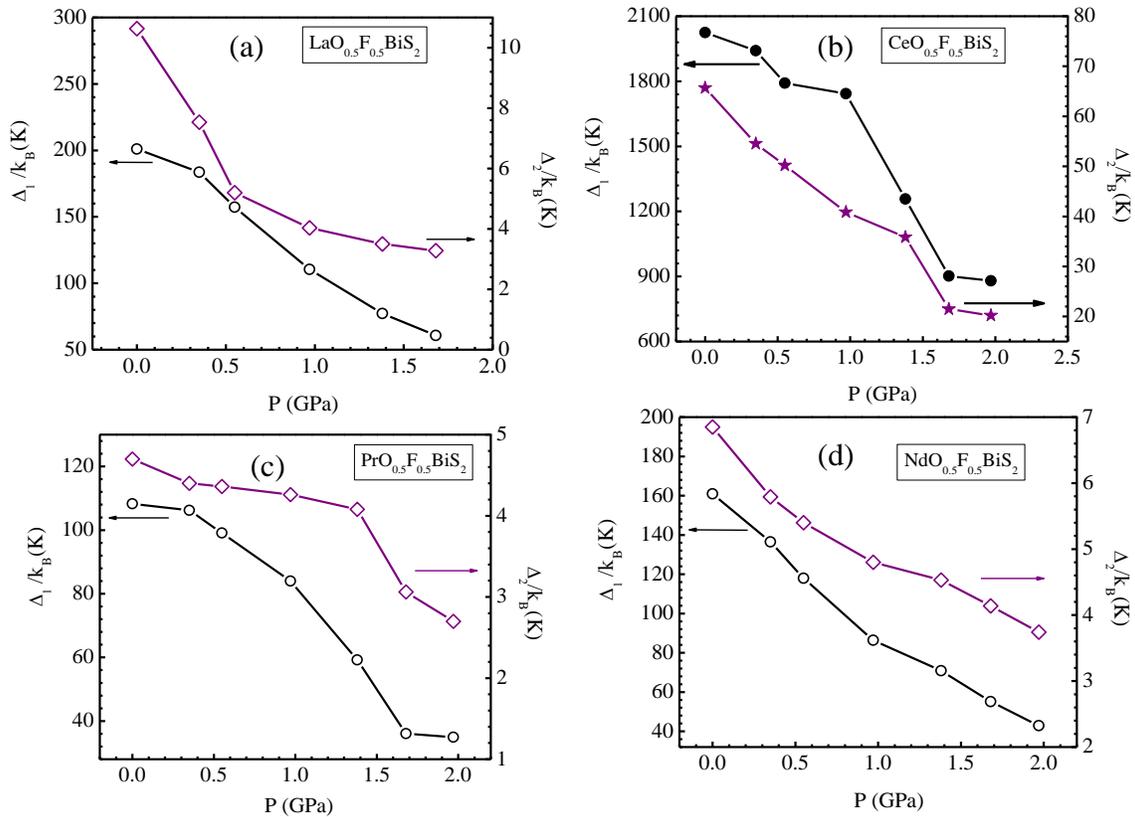

**Figure 5:** Energy gaps $\Delta_1$ and $\Delta_2$ as a function of pressure for (a) $LaO_{0.5}F_{0.5}BiS_2$, (b) $CeO_{0.5}F_{0.5}BiS_2$ (c) $PrO_{0.5}F_{0.5}BiS_2$ and (d) $NdO_{0.5}F_{0.5}BiS_2$ compounds.

With increases pressure energy gaps decreases; it may be due the applying pressure the bond angle and band structure change, which positively affect the electronic and magnetic correlations. The applying pressure can moderate the charge carrier's density at the Fermi surface in these $BiS_2$ based systems [54]. In some theoretical studies insulator-metal transition with superconducting transition at low temperature under pressure has been suggested for the $LaO_{0.5}F_{0.5}BiS_2$ compound [54]. In the present study, we observed that up to 1.97GPa pressure though there is no insulator to metal transition, but the resistivity is suppressed significantly and superconductivity onset transition temperature is increased for all the compounds. The interesting $BiS_2$-based superconductors need to be studied for further higher pressures.

Figure 6 (a-d) represent the temperature dependent resistivity under applied magnetic fields up to 50kOe of $LaO_{0.5}F_{0.5}BiS_2$, at 1.68GPa, $CeO_{0.5}F_{0.5}BiS_2$ at 1.97GPa, $PrO_{0.5}F_{0.5}BiS_2$ at 1.97GPa and $NdO_{0.5}F_{0.5}BiS_2$ at1.97GPa pressure. For all the compounds the $T_c$ decreases with increasing applied field as similar to the Type-II superconductor at ambient pressure. Interestingly, $LaO_{0.5}F_{0.5}BiS_2$ compound is more robust superconductor against magnetic field under pressure. The Anisotropic upper critical field has been reported for high-pressure synthesis$LaO_{0.5}F_{0.5}BiS_2$superconductor [101].

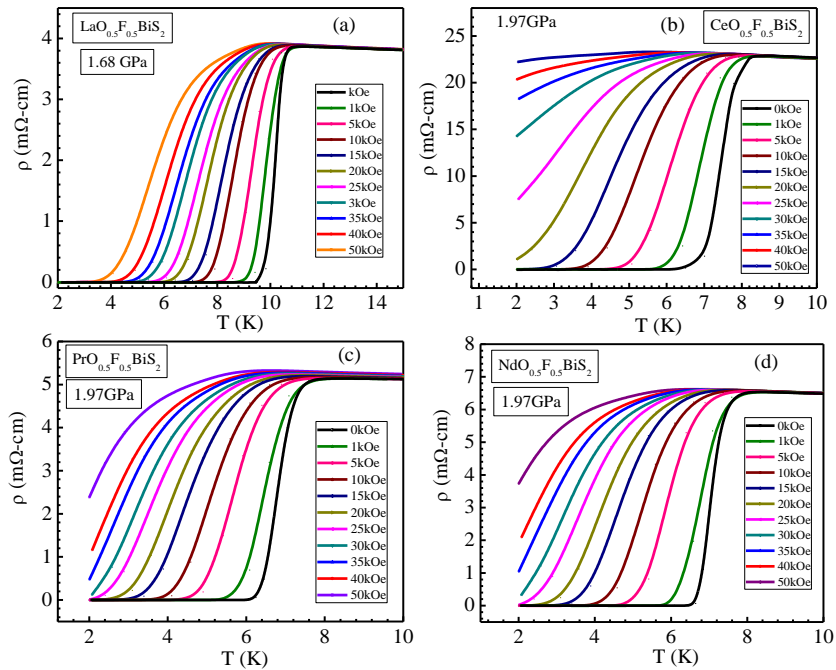

**Figure6:** $\rho(T)$ under magnetic fields for the (a) $LaO_{0.5}F_{0.5}BiS_2$ at 1.68GPa, (b) $CeO_{0.5}F_{0.5}BiS_2$ at 1.97GPa (c) $PrO_{0.5}F_{0.5}BiS_2$ at 1.97GPa, and (d) $NdO_{0.5}F_{0.5}BiS_2$ at1.97GPa

Relatively the superconductivity decreases with applied magnetic field $dT_c/dH$ is around 0.12K/kOe from absolute $T_c(\rho=0)$ criteria, which is near about the Fe Pnictides superconductors and more than the High $T_c$ superconductors. For the other compounds $CeO_{0.5}F_{0.5}BiS_2$, $PrO_{0.5}F_{0.5}BiS_2$ and $NdO_{0.5}F_{0.5}BiS_2$ $dT_c/dH$ from absolute $T_c(\rho=0)$ criteria is around 0.11K/kOe, 0.16K/kOe and 0.179K/kOe respectively.

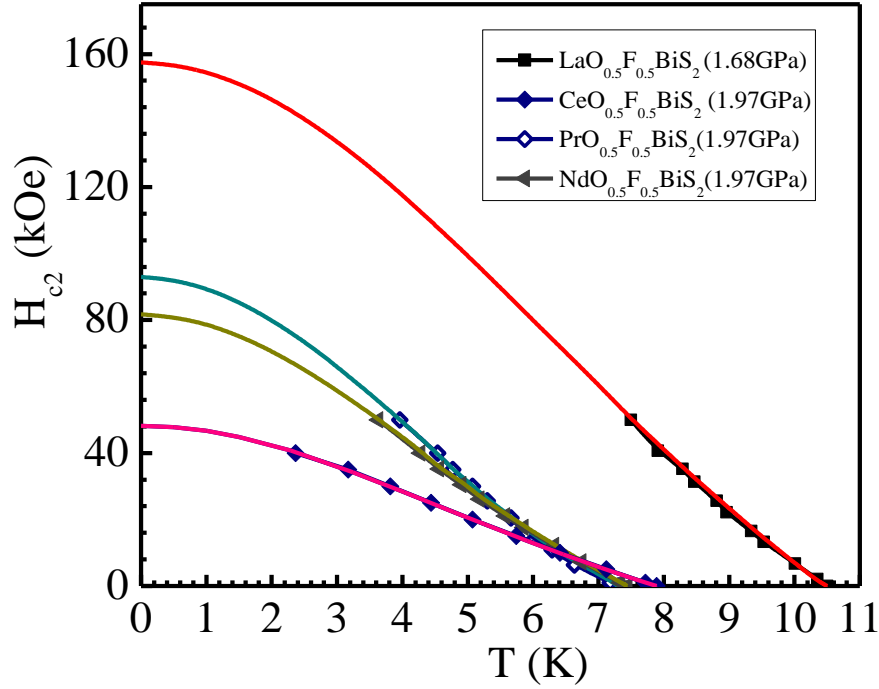

**Figure 7:** Upper critical field ($H_{c2}$) Vs T plots for the $LaO_{0.5}F_{0.5}BiS_2$ at 1.68GPa and $CeO_{0.5}F_{0.5}BiS_2$, $PrO_{0.5}F_{0.5}BiS_2$, $NdO_{0.5}F_{0.5}BiS_2$ at 1.97GPa hydrostatic pressures.

The upper critical field $H_{c2}$ of all the samples has been estimated by using the conventional one-band Werthamer–Helfand–Hohenberg (WHH) equation [102] for the 90% criterion of the normalized resistivity data, i.e., $H_{c2}(0)=-0.693T_c(dH_{c2}/dT)_{T=Tc}$. The estimated $H_{c2}(0)$ has been shown in the Figure 7 and values are 159kOe for $LaO_{0.5}F_{0.5}BiS_2$ compound, 48kOe for $CeO_{0.5}F_{0.5}BiS_2$, 88kOe for $PrO_{0.5}F_{0.5}BiS_2$, and 82kOe for $NdO_{0.5}F_{0.5}BiS_2$ compound.

**2.2 $Sr_{0.5}RE_{0.5}FBiS_2$(RE=La, Ce, Pr, NdandSm)**

**2.2.1 $Sr_{0.5}La_{0.5}FBiS_2$**

The temperature dependence of the electrical resistivity ρ below 250K down to 2K for $Sr_{0.5}La_{0.5}FBiS_2$ compound at various applied pressures of 0.35GPa-1.97GPa along with data without pressure are shown in Figure 8. The normal state resistivity behavior of the compound without applied pressure is clearly semiconducting down to superconductivity onset of around 2.2K, and superconductivity is seen with $T_c(\rho=0)$ at 2K. The normal state semiconducting behavior along with superconductivity at 2K is in agreement with reported data on $Sr_{0.5}La_{0.5}FBiS_2$ superconductor [24].

With applied pressure of 0.35GPa the normal state behavior of the resistivity changes to metallic and the same improves further for higher applied pressures upto 1.38GPa. For higher pressures of above 1.38GPa, the normal state behavior is though metallic, but with slightly less metallic slope. As far as normal state resistivity ($\rho^{20K}$), i.e., well above superconducting transition temperature onset is concerned the same is 10mΩ-cm for zero pressure, 0.128mΩ-cm for 0.35GPa, 0.106mΩ-cm for 1.38GPa, and 0.153mΩ-cm for 1.68GPa and 1.98GPa pressures. Clearly the normal state resistivity at 20K, when compared with ambient pressure sample, is decreased by more than an order of magnitude with 0.35GPa pressure, remains within same range for up to 1.38GPa and later increases slightly for 1.68GPa and 1.97GPa pressures. The decrease in normal state resistivity ($\rho^{20K}$), by more than an order of magnitude with 0.35GPa pressure along with the change of conduction process from semiconducting to metallic is surprising. It is possible that under pressure F-Sr/La-F bond angles along with the inter-atomic distances do change, which in turn affect the charge density at Fermi surface and hence a clear change in normal state electrical transport.

As far as superconductivity i.e., $T_c(\rho=0)$ is concerned the same can be seen clearly from the inset of Figure 8, which is zoomed part of same near superconducting transition state. The $T_c(\rho=0)$ for $Sr_{0.5}La_{0.5}FBiS_2$ sample without pressure is around 2K, which with applied pressure of 0.35GPa, and 0.55GPa is nearly the same but increases sharply to 8.6K for 0.97GPa. With further increase in pressure to 1.38GPa, 1.68GPa and 1.97GPa, the $T_c(\rho=0)$ is enhanced to above 10K. It is clear that superconducting transition temperature $T_c(\rho=0)$ remains nearly unchanged at around 2K for applied pressure till 0.55GPa and later increases sharply to around 10K for further higher pressure of 0.97GPa and remains nearly same (10K) till 1.97GPa. The fivefold increase in $T_c$, with applied pressure of just above 1GPa is surprising and calls for the possibility of unconventional superconductivity in these layered $BiS_2$ based superconductors. Such a huge increase in $T_c$ and decrease in normal state resistivity at small pressure suggest the structural phase transition in the $Sr_{0.5}La_{0.5}FBiS_2$ compound as similar $LaO_{0.5}F_{0.5}BiS_2$ compound [42,54].

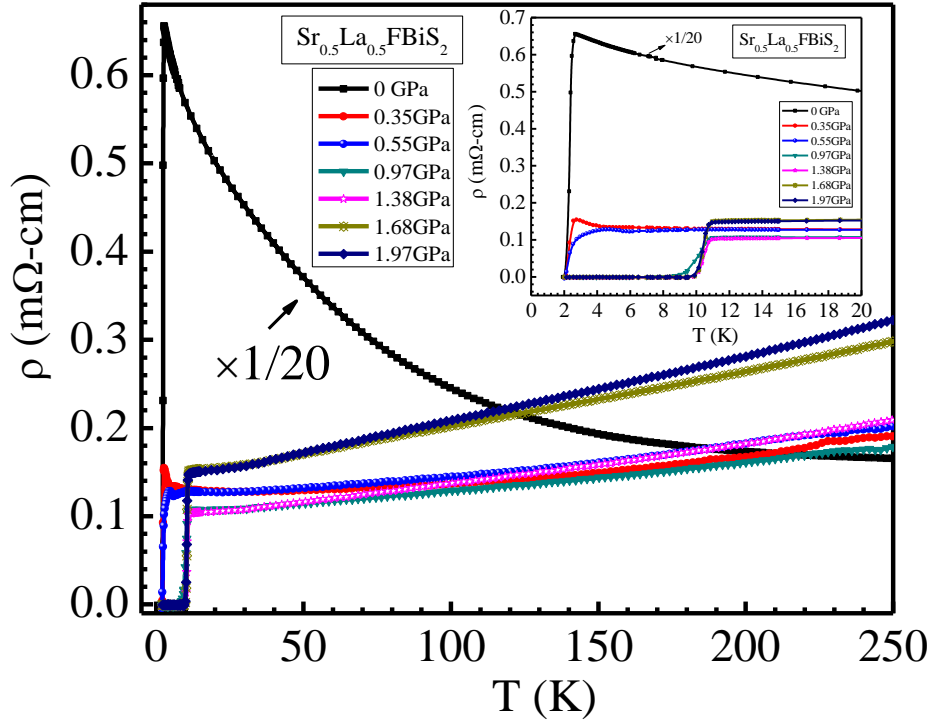

**Figure 8:** Resistivity versus temperature ($\rho$ Vs T) plots for $Sr_{0.5}La_{0.5}FBiS_2$ compound, at various applied pressures in the temperature range 250K-2K. Inset shows the same in the temperature range 20K-2.0K for $Sr_{0.5}La_{0.5}FBiS_2$ compound.

Summarizing, the results of Figure 8, It has been observed that with increasing pressure the normal state resistivity turns to the metallic behavior instead of the semiconducting along with a five-fold increase in superconducting transition temperature from 2K to 10K. This is though qualitatively same, but slightly different quantitatively from earlier reports on pressure dependent superconductivity of $BiS_2$ based superconducting $(La/Pr/Nd/Ce)O_{0.5}F_{0.5}BiS_2$ compounds [65-80]. In particular, though nearly fivefold increase was seen for superconducting transition temperature [77], the normal state conduction along with normal state resistivity is not improved so dramatically as in present case of $Sr_{0.5}La_{0.5}FBiS_2$ superconductor. Worth mentioning is the fact that this is first study on impact of hydrostatic pressure on superconductivity of $Sr_{0.5}La_{0.5}FBiS_2$ superconductor. Interestingly, $SrFBiS_2$ is proposed to be the parent phase for $BiS_2$ based superconductor family [24]. The coexistence of accompanied insulator to metal transition and fivefold increase in $T_c$ of $Sr_{0.5}La_{0.5}FBiS_2$ under moderate pressure of just above 1GPa warrants further detailed studies related to structural details under pressure and first principle calculations to explore possible strong electronic correlations in this newest class [42, 54] of BiS$_2$ based superconductors.

Figure 9 shows the temperature dependent resistivity of Sr$_{0.5}$La$_{0.5}$FBiS$_2$ superconductor at 0GPa and 1.97GPa pressure from 250K to 1.9K. The normal state resistance of Sr$_{0.5}$La$_{0.5}$FBiS$_2$ at 0GPa exhibits clearly the semiconducting behavior down to superconducting onset at 2.3K. At zero pressure the Resistivity versus temperature clearly shows the thermally activated behavior from 250K down to 2.3K. The thermal activation energy ($\Delta$) is obtained by fitting with thermal activation equation $\rho(T)=\rho_0\exp(\Delta/2k_BT)$ for the temperature range 250K to 90K. The activation energy is estimated about 10.2meV, which suggests a small energy gap in this compound. Similar semiconducting behavior of the resistivity has earlier been reported for LaO$_{0.5}$F$_{0.5}$BiS$_2$ [67,80].

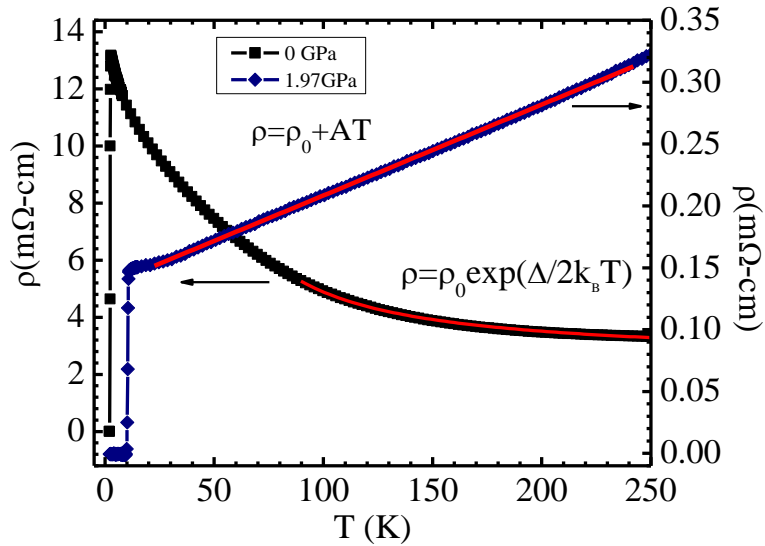

**Figure 9:** Resistivity versus temperature ($\rho$ Vs T) plots for Sr$_{0.5}$La$_{0.5}$FBiS$_2$ compound, at 0GPa and 1.97GPa pressure in the temperature range 250K-2K.

Interestingly, the charge density wave instability has been suggested recently [38] by the first-principles calculations, which may cause enhanced electron correlations in this system. The normal state resistivity shows metallic behavior as applied pressure increases. Here it can be clearly seen that the highest pressure 1.97GPa sample exhibit metallic conductivity with positive d$\rho$/dT along with a superconducting transition at below 10K. Red solid line shows the linearly fitted resistivity plot according to equation $\rho = \rho_0 + AT$, where $\rho_0$ is the residual resistivity and A is the slope of the graph.

### 2.2.2 $Sr_{0.5}Ce_{0.5}FBiS_2$

Figure 10 shows temperature dependent electrical resistivity $\rho(T)$ at various applied pressure 0.35-2.5GPa in the temperature range 2-300K for the $Sr_{0.5}Ce_{0.5}FBiS_2$ compound. The normal state resistivity shows the semiconducting behavior of the applied pressure 0.35GPa and is suppressed with increasing pressure up to 1.5GPa. The normal state resistivity completely changes from semiconducting to the metallic at the pressure 2GPa and 2.5GPa pressures.

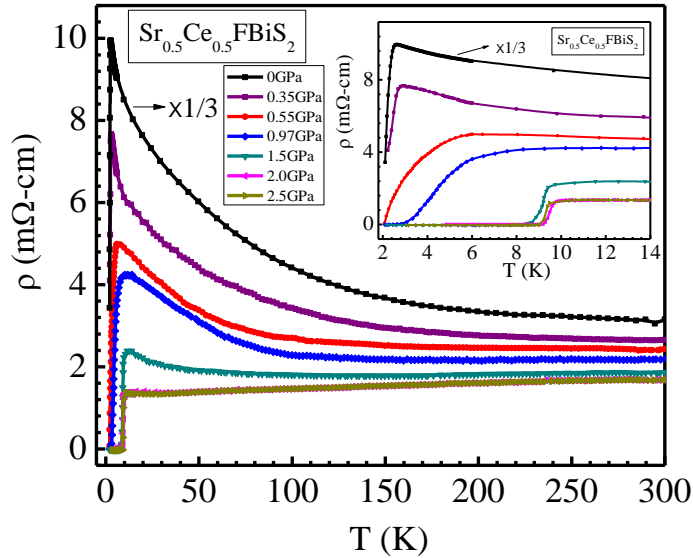

**Figure 10:** Resistivity versus temperature $\rho(T)$ plots for $Sr_{0.5}Ce_{0.5}FBiS_2$ compound at various applied pressures (0.35-2.5GPa) in the temperature range 300K-2.0K, inset of each shows the zoomed part of $\rho(T)$ plots near the superconducting transition temperature.

Inset of the Figure 10 shows the superconducting transition at various pressures. At pressure 0.35GPa, $T_c^{onset}$ appears at 2.8K it looks like ambient pressure data with reduced resistivity up to the 2K $T_c(\rho=0)$ not appears. The $T_c(\rho=0)$ above 2K can be seen at the pressure 0.55GPa with the $T_c^{onset}$ at 4K. The $T_c(\rho=0)$ increases from 2K to 3K at the applied pressure 0.97GPa with $T_c^{onset}$ at 6K, which having broadening in the transition. At the pressure 1.5GPa the superconductivity increases as $T_c^{onset}$ at 9.7K with $T_c(\rho=0)$ at 8.5K, for the 2.0GPa pressure $T_c^{onset}$ is 9.9K with the sharp transition. The superconducting transition temperature is nearly unchanged for the 2.5GPa pressure.

### 2.2.3 $Sr_{0.5}Pr_{0.5}FBiS_2$

The temperature dependent electrical resistivity at various applied pressure for the $Sr_{0.5}Pr_{0.5}FBiS_2$ is shown in the Figure 11. Also, in this compound semiconducting to metallic transition has been observed at the 2.0GPa. Inset is the zoomed part of Figure 11 in the temperature range 14-2K. Interestingly, at pressure 0.35GPa the compound shows $T_c(\rho=0)$ near 2K, and $T_c^{onset}$ is above 2.7K. At the 0.55GPa pressure $T_c^{onset}$ slightly increases 2.9K. The $T_c(\rho=0)$ above 2.5K and $T_c^{onset}$ is at 6K at the applied pressure 0.97GPa, the compound has large transition width at the this pressure. At the pressure 1.5GPa the superconductivity increases as $T_c^{onset}$ at 9.7K with $T_c(\rho=0)$ at 8.5K, for the 2.0GPa pressure $T_c^{onset}$ is 10K with the sharp transition, $T_c^{onset}$ is unaltered and $T_c(\rho=0)$ slightly decreases for the 2.5GPa pressure.

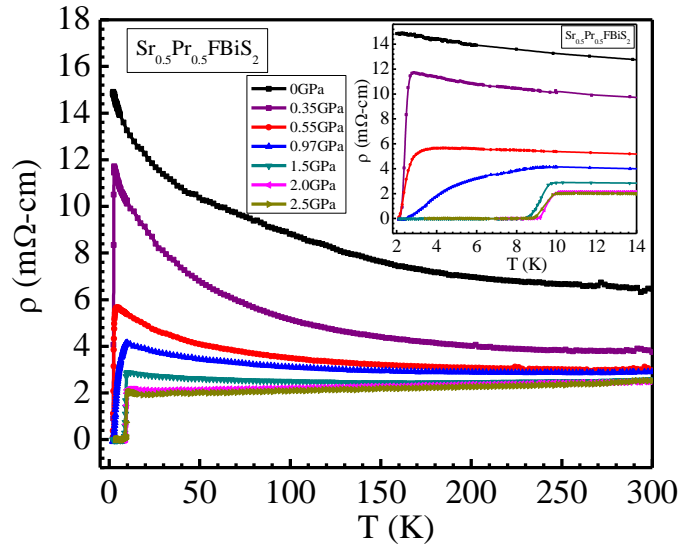

**Figure 11:** $\rho(T)$ plots for $Sr_{0.5}Pr_{0.5}FBiS_2$ compound at various applied pressures (0.35-2.5GPa) in the temperature range 300K-2K, inset of each shows the zoomed part of $\rho(T)$ plots near the superconducting transition temperature.

### 2.2.4 $Sr_{0.5}Nd_{0.5}FBiS_2$

Figure 12 shows $\rho(T)$ for the $Sr_{0.5}Nd_{0.5}FBiS_2$ compound in the temperature range 2-300K at various applied pressures of 0.35-2.5GPa. The semiconducting behavior of normal state resistivity also observed in the $Sr_{0.5}Nd_{0.5}FBiS_2$ compound from the applied pressure 0.35GPa to 1.5GPa with decreasing resistivity. The semiconducting to metallic transition has been observed at the 2GPa, which is followed for the further higher pressure of 2.5GPa. Inset is the zoomed part of Figure 12 in the temperature range 14-2K. At pressure 0.35GPa, the $T_c^{onset}$ appears at 2.5K, which is absent at the ambient pressure data down to 2K. The $T_c(\rho=0)$ above 2K is seen at 0.55GPa pressure, with the $T_c^{onset}$ at 3.5K. At the pressure 0.97GPa, the $T_c(\rho=0)$ slightly increases from 2K to 2.5K and $T_c^{onset}$ is at 5.8K, resulting in a broadened superconducting transition at this pressure. For the 1.5GPa pressure, the $T_c(\rho=0)$ is 4.5K and the $T_c^{onset}$ is at 9.25K, with a further broadened superconducting transition. The broadening in the superconducting transition reduces significantly at 2.0GPa pressure with $T_c(\rho=0)$ and $T_c^{onset}$ at 8.7K, and 9.5K respectively. For the further applied pressure of 2.5GPa the $T_c(\rho=0)$ is slightly increased to around 9K.

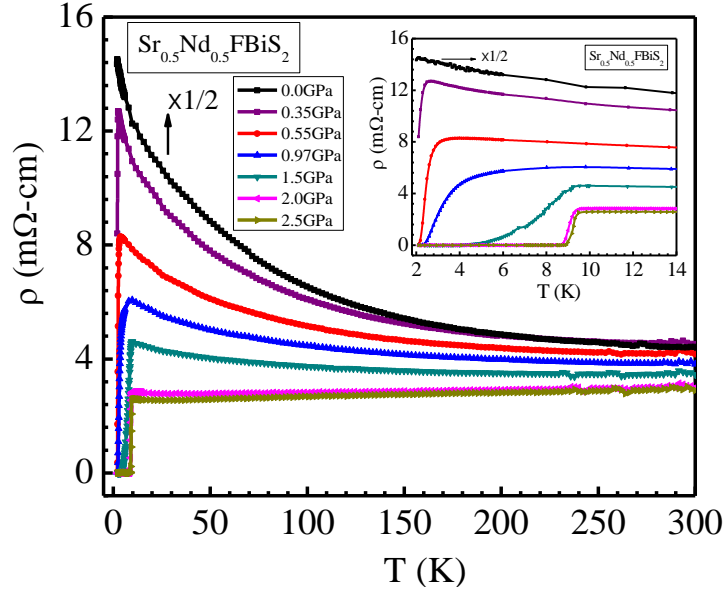

**Figure 12:** $\rho(T)$ plots for $Sr_{0.5}Nd_{0.5}FBiS_2$ compound, at various applied pressures (0.35-2.5GPa) in the temperature range 300K-2.0K, inset of each shows the zoomed part of $\rho(T)$ plots near the superconducting transition temperature.

### 2.2.5 $Sr_{0.5}Sm_{0.5}FBiS_2$

The $\rho(T)$ from 300K to 2K at various applied pressure for the $Sr_{0.5}Sm_{0.5}FBiS_2$ is shown in the Figure 13. The $Sr_{0.5}Sm_{0.5}FBiS_2$ compound shows the similar nature as for RE=La, Pr and Nd, i.e., semiconducting to metallic transition at the 2.0GPa. So, it can be said that the 2.0GPa is transition pressure for $Sr_{0.5}RE_{0.5}FBiS_2$ compound. Inset of Figure 13 is zoomed part in the superconducting region 14-2K. The $Sr_{0.5}Sm_{0.5}FBiS_2$ compound shows the $T_c^{onset}$ at 2.4K at 0.35GPa and 2.8K at at 0.55GPa, but not $T_c(\rho=0)$. At 0.97GPa pressure the $T_c^{onset}$ is near 5.6K, and $T_c(\rho=0)$ is above 2.2K. At 1.5GPa pressure the $T_c^{onset}$ slightly increases to 9.1K and the $T_c(\rho=0)$ is above 3.5K, with relatively larger transition width. For the further higher pressure of 2.0GPa the $T_c^{onset}$ is 9.4K and $T_c(\rho=0)$ at 8.0K. At 2.5GPa the superconducting transition is more sharper with slightly decreased $T_c^{onset}$ of 9.2K and $T_c(\rho=0)$ at 8.3K.

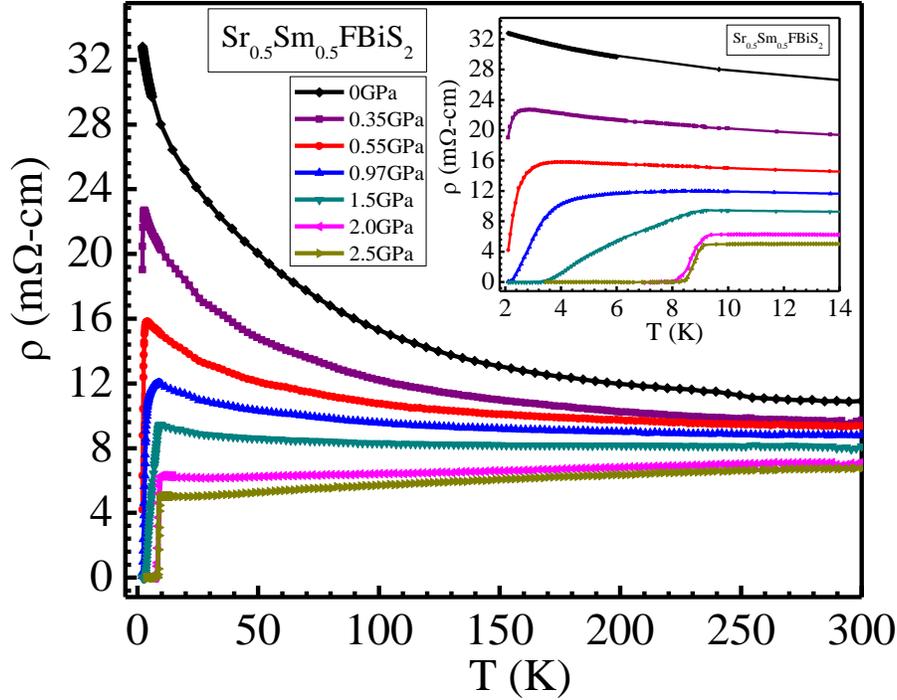

**Figure 13:** Resistivity versus temperature $\rho(T)$ plots for $Sr_{0.5}Sm_{0.5}FBiS_2$ compound at various applied pressures (0.35-2.5GPa) in the temperature range 300K-2.0K, inset of each shows the zoomed part of $\rho(T)$ plots near the superconducting transition temperature.

Figure 14 (a, b) shows the Pressure dependent superconducting transition temperature $T_c^{onset}$ and $T_c(\rho=0)$ for the $Sr_{0.5}RE_{0.5}FBiS_2$ (RE=La, Ce, Nd, Pr and Sm) compounds. At the ambient pressure $T_c^{onset}$ and $T_c(\rho=0)$ has been observed only for the $Sr_{0.5}La_{0.5}FBiS_2$ compound at 2.5 and 2.1K respectively. The $T_c^{onset}$ increases from 2.5K to 11K at just 0.97GPa and $T_c(\rho=0)$ increases from 2.1K to 8.5K. For the further higher pressure both the $T_c^{onset}$ and $T_c(\rho=0)$ saturate at 11K and 9.4K respectively. The $T_c^{onset}$ observed at 2.8K for the $Sr_{0.5}Ce_{0.5}FBiS_2$ compound $T_c(\rho=0)$ absent down to 2K and the $T_c(\rho=0)$ appears at the applied pressure 0.55GPa with the $T_c^{onset}$ at 4K. The jump in both $T_c^{onset}$ and $T_c(\rho=0)$ observed at the pressure 1.5GPa increases as $T_c^{onset}$ at 9.7K with $T_c(\rho=0)$ at 8.5K, for the 2.0GPa pressure $T_c^{onset}$ is 9.9K with the sharp transition. The $Sr_{0.5}Pr_{0.5}FBiS_2$ compound at pressure 0.35GPa shows $T_c(\rho=0)$ near 2K, and $T_c^{onset}$ is above 2.7K, which slightly increases $T_c(\rho=0)$ =2.5K and $T_c^{onset}$ = 2.9K at 0.55GPa pressure. The superconductivity enhance as $T_c^{onset}$ at 9.7K with $T_c(\rho=0)$ at 8.5K for the pressure 1.5GPa. The $T_c^{onset}$ is 10K with the $T_c(\rho=0)$ 8.8K for the 2.0GPa pressure, $T_c^{onset}$ is unchanged and $T_c(\rho=0)$ slightly decreases for the 2.5GPa pressure. For the $Sr_{0.5}Nd_{0.5}FBiS_2$ compound the $T_c^{onset}$ increases at 9K at pressure 1.5GPa with $T_c(\rho=0)$ at 4K. The superconducting transition being sharp at the pressure 2.0GPa with $T_c^{onset}$ = 9.5K and $T_c(\rho=0)$ at 8.7K. For the $Sr_{0.5}Sm_{0.5}FBiS_2$ compound $T_c(\rho=0)$ at 2.2K obtained at 0.97GPa pressure with the $T_c^{onset}$ = 5.6K. At 1.5GPa pressure the $T_c^{onset}$ increases to 9.1K and the $T_c(\rho=0)$ is above 3.5K, for pressure 2.0GPa the $T_c^{onset}$ is 9.4K and $T_c(\rho=0)$ at 8.0K. The $T_c^{onset}$ slightly decreased to 9.2K with $T_c(\rho=0)$ at 8.3K at the 2.5GPa pressure. Further it seems that 1.5GPa is the transition pressure ($P_t$) for sharp increase in $T_c^{onset}$ of $Sr_{0.5}RE_{0.5}FBiS_2$ (RE=Ce, Nd, Pr and Sm) compounds. If we further increase the pressure (>1.5GPa), the $T^{onset}$ is almost saturated, but the superconducting transition width is sharper in this pressure range with increased $T_c(\rho=0)$.

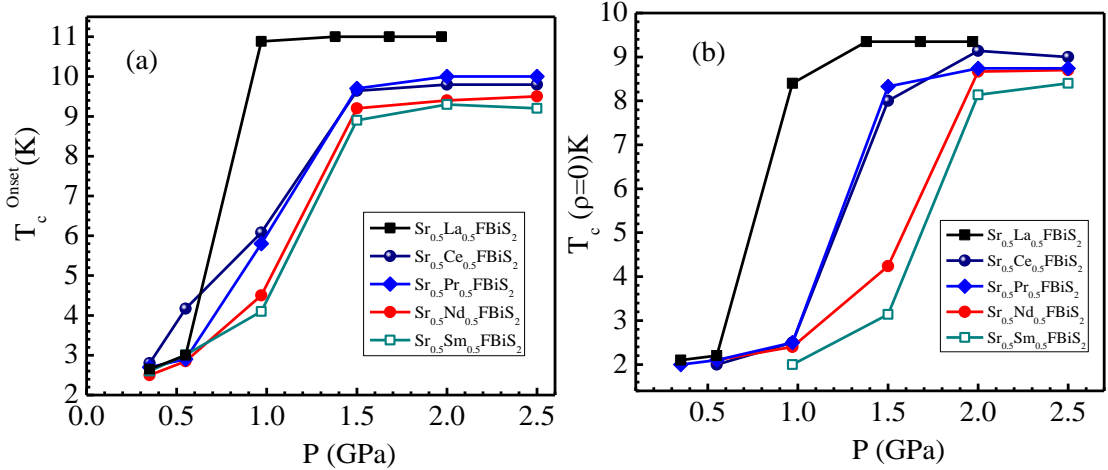

**Figure 14:**(a, b)$T_c^{onset}$ and Tc ($\rho=0$) vs applied Pressure for the $Sr_{0.5}RE_{0.5}FBiS_2$ (La, Ce, Pr, Nd and Sm) compounds.

**Magneto-transport properties**

Figure 15(a-e) shows the $\rho(T)$ under applied magnetic fields (0-70kOe) for the $Sr_{0.5}RE_{0.5}FBiS_2$ (RE=La, Ce, Nd, Pr and Sm) compounds under the hydrostatic pressure of 1.97GPa for $Sr_{0.5}La_{0.5}FBiS_2$ and 2.5GPa for $Sr_{0.5}RE_{0.5}FBiS_2$ (RE= Ce, Nd, Pr and Sm). It can be clearly seen that the $T_c^{onset}$ decreases less as compared to $T_c(\rho=0)$ with the increasing magnetic field. In these compounds, under magnetic field the transition width is broadened. The upper critical field $H_{c2}$ has been estimated from the resistivity drop 90% the normal state resistance $\rho_n(T,H)$ at $T_c^{onset}$ in applied magnetic fields for all the samples. The upper critical field at absolute zero temperature $H_{c2}(0)$ can be determined by using the conventional one-band

Werthamer–Helfand–Hohenberg (*WHH*) equation [102], i.e., $H_{c2}(0) = -0.693 T_c (dH_{c2}/dT)_{T=T_c}$. The estimated $H_{c2}(0)$ are 200kOe, 138kOe, 116kOe, 135kOe and 145kOe for $Sr_{0.5}La_{0.5}FBiS_2$ at pressure 1.97GPa, $Sr_{0.5}Ce_{0.5}FBiS_2$ at pressure 2.5GPa, $Sr_{0.5}Nd_{0.5}FBiS_2$ at pressure 2.5GPa, $Sr_{0.5}Pr_{0.5}FBiS_2$ at pressure 2.5GPa and $Sr_{0.5}Sm_{0.5}FBiS_2$ at pressure 2.5GPa respectively.

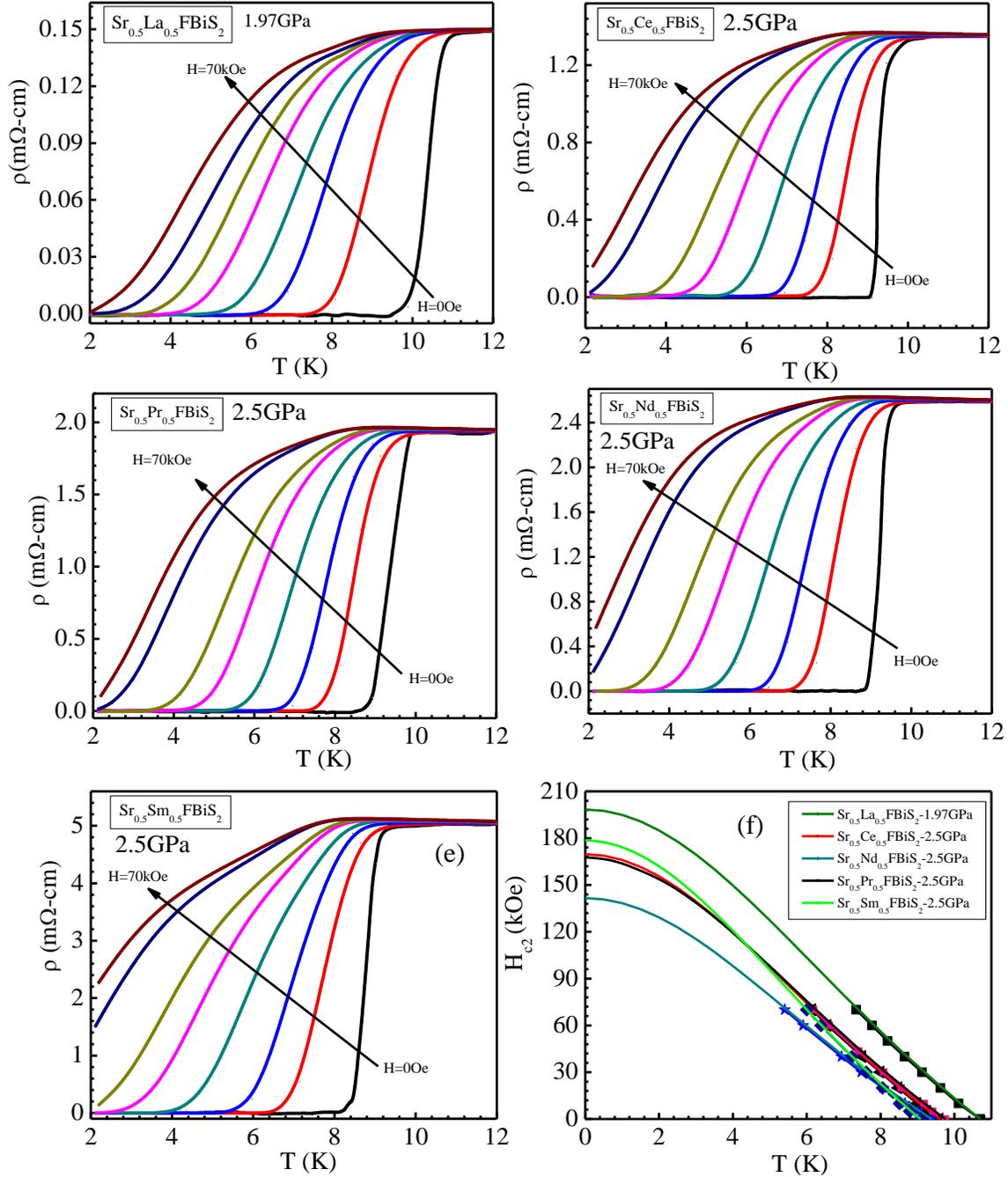

**Figure 15: (a-e)** Temperature dependence of the Resistivity $\rho(T)$ under magnetic fields for the $Sr_{0.5}RE_{0.5}FBiS_2$ compounds under hydrostatic pressure. **(f)** The upper critical field $H_{c2}$ taken from 90%, Resistivity criterion $\rho(T)$ for the $Sr_{0.5}RE_{0.5}FBiS_2$ under hydrostatic pressure.

As shown in Figure 15(f), the solid lines are the extrapolation to the Ginzburg–Landau equation $H_{c2}(T)=H_{c2}(0)(1-t^2/1+t^2)$, where $t=T/T_c$ is the reduced temperature which gives values slightly higher than that by WHH approach. These upper critical field values for all samples are close to but within the Pauli paramagnetic limit i.e. $H_p=1.84T_c$.

**2.3 SrFBiS$_2$**

We observed an insulating normal state to metallic for compounds $Sr_{0.5}RE_{0.5}FBiS_2$ (RE=Ce, Nd, Pr and Sm) under hydrostatic pressure above the 1GPa, which is accompanied with increased superconductivity up to 10K [77]. We measure the temperature dependent electrical resistivity from 300K down to 2K under applied the hydrostatic pressure (0-2.5GPa) for SrFBiS$_2$, i.e., parent phase for $Sr_{0.5}RE_{0.5}FBiS_2$ compounds.

Figure 16 shows the temperature dependence of electrical resistivity under the different hydrostatic pressure (0-2.5GPa) for SrFBiS$_2$ down to 2K. A clear semiconducting behavior can be observed in the temperature range 300-2K as temperature coefficient dρ/dT<0 is negative [96]. With increasing applied pressure the electrical resistivity significantly decrease. At the applied pressures of 2.17 and 2.5GPa the electrical resistivity becomes almost temperature independent. The inset of the Figure 63 shows the zoomed portion of same in the temperature range 8K to 2K. There is no indication of superconducting transition down to 2K even after application of up to 2.5GPa pressure. The suppression of resistivity in SrFBiS$_2$ under hydrostatic pressure is similar to the BiS$_2$-based layered superconductors $LaO_{0.5}F_{0.5}BiS_2$ and $Sr_{0.5}La_{0.5}FBiS_2$, but superconductivity is not observed.

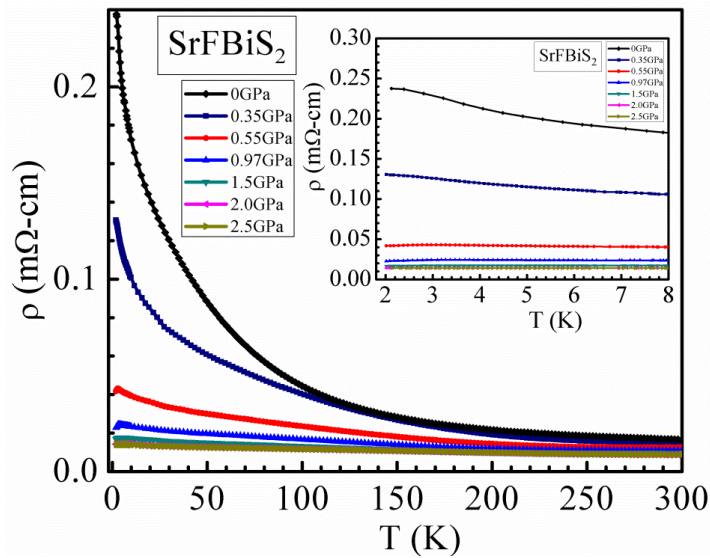

**Figure 16:** Temperature (T) dependences of resistivity (ρ) under hydrostatic pressure of compound SrFBiS$_2$ compound and inset is the zoom part from 2-8K of the same.

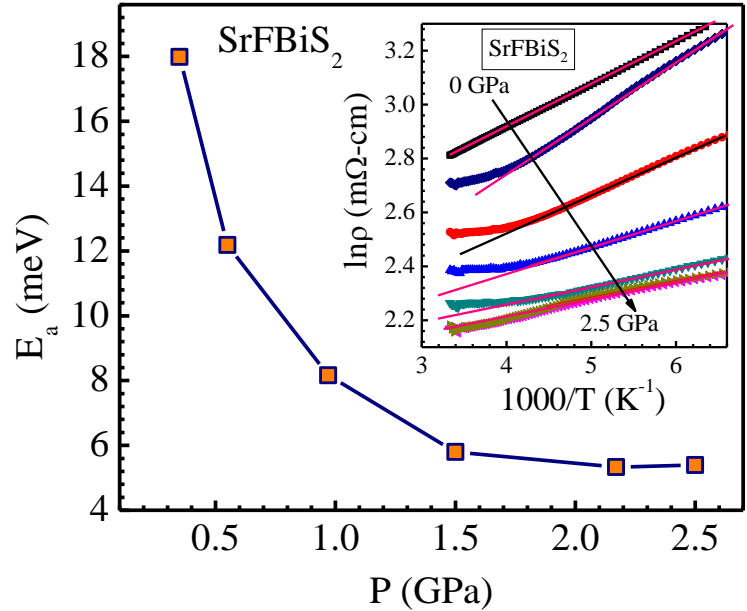

**Figure 17:** Pressure dependent activation (E$_g$) evaluated from thermally activation equation where solid line is a guide to eyes. Inset is the lnρ Vs 1/T for the different pressures (0-2.5GPa) of SrFBiS$_2$ compound where solid lines are the extrapolation to the fitted portion.

Figure 17 presents the thermal activation energy (E$_a$) as a function of applied hydrostatic pressure as obtained from fitting the resistivity to the equation $\rho(T)=\rho_0 \exp(E_a/k_B T)$ in the temperature range 300-100K. The thermal activation energy drops exponentially with pressure. Inset of the Figure 17 is the ln(ρ) Vs 1/T to evaluate the activation energy of SrFBiS$_2$ from 0-2.5GPa pressure. It is interesting to note that the value of $\rho_0$ under different hydrostatic pressures remains almost same close to 7.2mΩ-cm. This observation suggests that at highest temperature the resistivity of sample is almost constant even under different hydrostatic pressure. The activation energy decreases with increasing pressure and saturates close to 5.3meV. This suggests that the Fermi energy moves close to conduction band by increasing pressure while highest temperature resistivity remains constant. The detailed spectroscopic and theoretical investigations needed under pressure to validate this point.

## 3. Experimental details

The BiS$_2$-based bulk polycrystalline REO$_{1-x}$F$_x$BiS$_2$ (RE=La, Ce, Pr and Nd) compounds were synthesized through the solid state reaction root via vacuum encapsulation. The high purity (~99.9%) rare earth metal (La, Ce, Pr and Nd), rare earth oxide ($La_2O_3$, $Nd_2O_3$, $Pr_6O_{11}$, & $CeO_2$,), Rare earth Fluoride, Bismuth, and Sulfur chemicals are taken in the stoichiometric ratio and ground thoroughly in the glove box (MBRAUN Lab star) filled with Argon gas of high purity (4N). The mixed powder subsequently palletized and Vacuum sealed ($10^{-4}$ mbar) in quartz tube. The vacuum encapsulated samples were heated in the box furnace at temperature 800°C with typical heating rate of 2°C/min for 12h. Another $BiS_2$-based $Sr_{0.5}La_{0.5}FBiS_2$ compound was synthesized by using high purity La, $SrF_2$, Bi and S. The process is similar up to the vacuum encapsulation step. The box furnace was used to sinter the sample at 650°C for 12h with the typical heating rate of 2°C/min. The sintered samples were subsequently cooled down slowly with the cooling rate 5°C/min to room temperature.

The electrical transport measurements were performed on Physical Property Measurements System (*PPMS*-14T, *Quantum Design*) as a function of both temperature and applied magnetic field. The pressure dependent resistivity measurements were performed by usingHPC-33 Piston type pressure cell with Quantum design DC resistivity Option on *PPMS*-14T, *Quantum Design*. Hydrostatic pressures were generated by a BeCu/NiCrAl clamped piston-cylinder cell. The sample was immersed in a fluid pressure transmitting medium of Fluorinert (FC70:FC77=1:1) in a Teflon cell. The resistivity measured by using four point probe method. Annealed Pt leads were affixed to gold-sputtered contact surfaces on each sample with silver epoxy in a standard four-wire configuration. The pressure at low temperature was estimated from the superconducting transition temperature of Pb.

## 4. Conclusion

Impact of external pressure on the Superconductors play an important for the discovery of new superconductors. In this regards, the high $T_c$ Cuprates and Fe-based superconductors had already set the superconductivity research in the forefront of condensed matter physics. Similarly, the newly discovered $BiS_2$ based compounds responded in a big way towards the impact of hydrostatic pressure on their superconductivity. In this review the electrical resistivity under hydrostatic pressure for $BiS_2$ base new superconductors is presented in a consolidated way at one place.

## 5. References

[1] Y. Mizuguchi, H. Fujihisa, Y. Gotoh, K. Suzuki, H. Usui, K. Kuroki, S. Demura, Y. Takano, H. Izawa, O. Miura, Phys. Rev. B **86**, 220510(R) (2012).

[2] S. K. Singh, A. Kumar, B. Gahtori, Shruti; G. Sharma, S. Patnaik, V. P. S. Awana, J. Am. Chem. Soc., **134,** 16504 (2012).

[3] S. Li, H. Yang, D. L. Fang, Z. Y. Wang, J. Tao, X. X. Ding, H.-H. Wen, Sci. China-Phys. Mech. Astron. **56**, 2019 (2013).

[4] S. G. Tan, L. J. Li, Y. Liu, P. Tong, B. C. Zhao, W. J. Lu, Y. P. Sun Physica C, **483**, 94 (2012).

[5] HiroshiTakatsu, YoshikazuMizuguchi, HirokiIzawa, Osuke Miura, HiroakiKadowaki, J. Phys. Soc. Jpn. **81** 125002 (2012).

[6] J. Shao, Z. Liu, X. Yao, L. Pi, S. Tan, C. Zhang, Y. Zhang, Physica Status Solidi (RRL) **8**, 845 (2014).

[7] Y. Mizuguchi, S. Demura, K. Deguchi, Y. Takano, H. Fujihisa, Y. Gotoh, H. Izawa, O. Miura, J. Phys. Soc. Jpn. **81**, 114725 (2012).

[8] V.P.S. Awana, A. Kumar, R. Jha, S. K. Singh, A. Pal, Shruti, J. Saha, S. Patnaik, Solid State Communications **157** 21 (2013).

[9] S. Demura, Y. Mizuguchi, K.Deguchi, H. Okazaki, H. Hara, T. Watanabe, S. J. Denholme, M. Fujioka, T. Ozaki, H. Fujihisa, Y. Gotoh, O. Miura, T. Yamaguchi, H. Takeya, and Y. Takano, J. Phys. Soc. Jpn. **82**, 033708 (2013).

[10] J. Xing, S. Li, X. Ding, H. Yang, and H. H. Wen, Phys. Rev. B **86**, 214518 (2012).

[11] R. Jha, A. Kumar, S. K. Singh, V. P. S. Awana, J Supercond Nov Magn. **26**, 499 (2013).

[12] R. Jha, A. Kumar, S. K. Singh, and V. P. S. Awana, J. App. Phy. **113**, 056102 (2013).

[13] D. Yazici, K. Huang, B.D. White, A.H. Chang, A.J. Friedman and M.B. Maple, Philosophical Magazine, **93**, 673 (2013).

[14] K. Deguchi, Y. Mizuguchi, S. Demura, H. Hara, T. Watanabe, S. J. Denholme, M. Fujioka, H. Okazaki, T. Ozaki, H. Takeya, T. Yamaguchi, O. Miura and Y. Takano, , Euro Phys. Lett. **101** 17004 (2013).

[15] A. Miura, M. Nagao, T. Takei, S. Watauchi, I. Tanaka, N. Kumada, J. Solid State Chem. 212, 213 (2014).

[16] A. Omachi, J. Kajitani, T. Hiroi, O. Miura, Y. Mizuguchi, J. Appl. Phys. **115**, 083909 (2014).

[17] K. Deguchi, Y. Takano, Y. Mizuguchi, Sci. Technol. Adv. Mater. **13**, 054303 (2013).

[18] J.C. Bednorz and K.A. Müller, ZeitschriftfürPhysik B. **64,** 189 (1986).

[19] Y. Kamihara, H. Hiramatsu, M. Hirano, R. Kawamura, H. Yanagi, T. Kamiya, and H. Hosono, J. Am. Chem. Soc., **128,** 10012 (2006).

[20] R. Ceolin, N. Rodier, ActaCrystallogr. B, **32**, 1476 (1976).

[21] V. S. Tanryverdiev, O. M. Aliev, I. Aliev, Inorg. Mater, **31**, 1361 (1995).

[22] D. Yazici, K. Huang, B. D. White, I. Jeon, V. W. Burnett, A. J. Friedman, I. K. Lum, M. Nallaiyan, S. Spagna, and M. B. Maple, Phys. Rev. B, **87**, 174512 (2013).

[23] X. Lin, X. Ni, B. Chen, X. Xu, X. Yang, J. Dai, Y. Li, X. Yang, Y. Luo, Q. Tao, G. Cao, and Z. Xu, Phys. Rev. B, **87**, 020504(R) (2013).

[24] H. Lei, K. Wang, M. Abeykoon, E. S. Bozin, and C. Petrovic, Inorg. Chem. **52**, 10685 (2013).

[25] H. Chen, G. Zhang, T. Hu, G. Mu, W. Li, F. Huang, X. Xie, and M. Jiang, Inorg. Chem., **53**, 9 (2014).

[26] T. Hiroi, J. Kajitani, A. Omachi, O. Miura, Y. Mizuguchi, arXiv:1404.6359

[27] I. Jeon, D. Yazici, B. D. White, A. J. Friedman, and M. B. Maple Phys. Rev. B **90**, 054510 (2014).

[28] Y. Fang, D. Yazici, B. D. White, M. B. Maple, arXiv:1410.0061

[29] G. S. Thakur, G. K. Selvan, Z. Haque, L. C. Gupta, S. L. Samal, S. Arumugam and A. K. Ganguli, arXiv:1410.0751.

[30] J. Kajitani, T. Hiroi, A. Omachi, O. Miura, Y. Mizuguchi, arXiv:1404.6361.

[31] J. Kajitani, A. Omachi, T. Hiroi, O. Miura, Y. MizuguchiPhysica C, in printing (ISS2013 proceedings).

[32] Y. Mizuguchi, A. Omachi, Y. Goto, Y. Kamihara, M. Matoba, T. Hiroi, J. Kajitani, O. Miura, arXiv:1409.2189

[33] Y. Mizuguchi, arXiv:1311.4272.

[34] W. A. Phelan, D. C. Wallace, K. E. Arpino, J. R. Neilson, K. J. Livi, C. R. Seabourne, A. J. Scott, and T. M. McQueen., J. Am. Chem. Soc., **135**, 5372 (2013).

[35] Y. Yu, J. Shao, S. Tan, C. Zhang, and Y. Zhang, J. Phys. Soc. Jpn **82,** 034718 (2013).

[36] P. Srivatsava, Shruti, S. Patnaik, Supercond. Sci. Technol. **27,** 055001 (2014).

[37] Shruti, P. Srivastava and S. Patnaik, J. Phys. Condens. Matter **25**, 312202 (2013).

[38] H. Usui, K. Suzuki, and K. Kuroki, Phys. Rev. B **86**, 220501(R) (2012).

[39] Y. Yang, W. S. Wang, Y. Y. Xiang, Z. Z. Li, and Q. H. Wang, Phys. Rev. B **88**, 094519 (2013).

[40] T. Zhou, Z. D. Wang, J Supercond Nov Magn, **26**, 2735 (2013).

[41] X. Wan, H. C. Ding, S. Y. Savrasov, and C. G. Duan, Phys. Rev. B, **87**, 115124 (2013).

[42] T. Yildirim, Phys. Rev. B**87**, 020506(R) (2013).

[43] B. Liu, S. Feng, EPL **106**, 17003 (2014).

[44] S. L. Liu, J. Supercond. Novel Mag. **26**, 3411 (2013).

[45] I. R. Shein, A. L. Ivanovskii, JETP Letters, **96**, 769 (2012).

[46] Y. Liang, X. Wu, W.-F. Tsai, J. Hu, Frontiers of Physics **9**, 194 (2014).

[47] K. Suzuki, H. Usui, K. Kuroki, Physics Procedia**45**, 21 (2013).

[48] G. Martins, A. Moreo, E. Dagotto, Physical Review B **87**, 081102(R) (2013).

[49] Y. Gao, T. Zhou, H. Huang, P. Tong, Q.-H. Wang, Phys. Rev. B **90**, 054518 (2014).

[50] X. Wu, J. Yuan, Y. Liang, H. Fan, J. Hu, EPL**108**, 27006 (2014).

[51] G. Lamura, T. Shiroka, P. Bonfa, S. Sanna, R. De Renzi, C. Baines, H. Luetkens, J. Kajitani, Y. Mizuguchi, O. Miura, K. Deguchi, S. Demura, Y. Takano, and M. Putti, Phys. Rev. B **88**, 180509(R) (2013).

[52] B. Li, Z. W. Xing and G. Q. Huang, EPL, **101**, 47002 (2013).

[53] C. Morice, E. Artacho, S. E. Dutton, D. Molnar, H.-J. Kim, S. S. Saxena, arXiv:1312.2615.

[54] C. Morice, E. Artacho, S. E. Dutton, D. Molnar, H. J. Kim, and S. S. Saxena, Phys. Rev B **92**, 041113(R) (2015).

[55] Yi Gao, arXiv:1304.2102.

[56] J. Lee, M. B. Stone, A. Huq, T. Yildirim, G. Ehlers, Y. Mizuguchi, O. Miura, Y. Takano, K. Deguchi, S. Demura, and S.-H. Lee, Phys. Rev. B,**87**, 205134 (2013).

[57] S. Nagira, J. Sonoyama, T. Wakita, M. Sunagawa, Y. Izumi, T. Muro, H. Kumigashira, M. Oshima, K. Deguchi, H. Okazaki, Y. Takano, O. Miura, Y. Mizuguchi, K. Suzuki, H. Usui, K. Kuroki, K. Okada, Y. Muraoka, T. Yokoya, J. Phys. Soc. Jpn. **83** 033703 (2014).

[58] L. K. Zeng, X. B. Wang, J. Ma, P. Richard, S. M. Nie, H. M. Weng, N. L. Wang, Z. Wang, T. Qian, H. Ding, Phys. Rev. B **90**, 054512 (2014).

[59] Z. R. Ye, H. F. Yang, D. W. Shen, J. Jiang, X. H. Niu, D. L. Feng, Y. P. Du, X. G. Wan, J. Z. Liu, X. Y. Zhu, H. H. Wen, M. H. Jiang, Phys. Rev. B **90**, 045116 (2014).

[60] X. B. Wang, S. M. Nie, H. P. Wang, P. Zheng, P. Wang, T. Dong, H. M. Weng, N. L. Wang Phys. Rev. B **90**, 054507 (2014).

[61] T. Machida, Y. Fujisawa, M. Nagao, S. Demura, K. Deguchi, Y. Mizuguchi, Y. Takano, H. Sakata J. Phys. Soc. Jpn 83, 113701 (2014).

[62] S. F. Wu, P. Richard, X. B. Wang, C. S. Lian, S. M. Nie, J. T. Wang, N. L. Wang, H. Ding Phys. Rev. B **90**, 054519 (2014).

[63] H.-F. Zhai, Z.-T. Tang, H. Jiang, K. Xu, K. Zhang, P. Zhang, J.-K. Bao, Y.-L. Sun, W.-H. Jiao, I. Nowik, I. Felner, Y.-K. Li, X.-F. Xu, Q. Tao, C.-M. Feng, Z.-A. Xu, G.-H. Cao, Phys. Rev. B **90**, 064518 (2014).

[64] H.-F. Zhai, P. Zhang, S.-Q Wu, C.-Y. He, Z.-T. Tang, H. Jiang, Y.-L. Sun, J.-K. Bao, I. Nowik, I. Felner, Y.-W. Zeng, Y.-K. Li, X.-F. Xu, Q. Tao, Z.-A. Xu, and G.-H. Cao,J. Am. Chem. Soc., **136**, 15386 (2014).

[65] H. Kotegawa, Y. Tomita, H. Tou, H. Izawa, Y. Mizuguchi, O. Miura, S. Demura, K. Deguchi, and Y. Takano,. J. Phys. Soc. Jpn **81** 103702 (2012).

[66] G. KalaiSelvan, M. Kanagaraj, S. EsakkiMuthu, R. Jha, V. P. S. Awana, and S. Arumugam, Phys. Status Solidi RRL,**7**, 510 (2013).

[67] C. T. Wolowiec, D. Yazici, B. D. White, K. Huang, and M. B. Maple, Phys. Rev. B **88**, 064503 (2013).

[68] C. T. Wolowiec, B. D. White, I. Jeon, D. Yazici, K. Huang and M. B. Maple, J. Phys. Cond. Matt. **25**,422201 (2013).

[69] T. Tomita, M. Ebata, H. Soeda, H. Takahashi, H. Fujihisa, Y. Gotoh, Y. Mizuguchi, H. Izawa, O. Miura, S. Demura, K. Deguchi, Y. Takano, J. Phys. Soc. Jpn. **83**, 063704 (2014).

[70] Y. Mizuguchi, T. Hiroi, J. Kajitani, H. Takatsu, H. Kadowaki, O. Miura J. Phys. Soc. Jpn. **83,** 053704 (2014).

[71] R. Jha, H. Kishan and V.P.S. Awana, Solid State Communications, **194,** 6(2014).

[72] M. Fujioka, M. Nagao, S. J. Denholme, M. Tanaka, H. Takeya, T. Yamaguchi, Y. Takano, Appl. Phys. Lett., **105**, 052601 (2014).

[73] J. Kajitani, T. Hiroi, A. Omachi, O. Miura, and Y. Mizuguchi, J. Phys. Soc. Jpn. **84,** 044712 (2015).

[74] J. Kajitani, K. Deguchi, A. Omachi, T. Hiroi, Y. Takano, H. Takatsu, H. Kadowaki, O. Miura, Y. Mizuguchi, Solid State Commun. **181**, 1 (2014).

[75] J. Kajitani, K. Deguchi, T. Hiroi, A. Omachi, S. Demura, Y. Takano, O. Miura, Y. Mizuguchi, J. Phys. Soc. Jpn. **83**, 065002 (2014).

[76] I. Pallecchi, G. Lamura, M. Putti, J. Kajitani, Y. Mizuguchi, O. Miura, S. Demura, K. Deguchi, Y. Takano, Phys. Rev. B **89,** 214513 (2014).

[77] R. Jha, Brajesh Tiwari and V.P.S. Awana. J. Phys. Soc. Jpn. **83**, 063707 (2014).

[78] R. Jha, B. Tiwari, V.P.S. Awana, J. Phys. Soc. Jpn, **83**, 105001 (2014).

[79] R. Jha, B. Tiwari, V.P.S. Awana, J. Appl. Phys. **117,** 013901 (2015)

[80] R. Jha, H. Kishan, V. P. S. Awana, J. Phys. Chem. Sol. **84,** 17 (2015).

[81] C. W. Chu, T. F. Smith, and W. E. Gardner, Phys. Rev. Letters **20**, 198 (1968).

[82] C. I. Sathish, H. L. Feng, Y. Shi, and K. Yamaura, J. Phys. Soc. Jpn **82** 074703 (2013).

[83] S.G. Tan, P. Tong, Y. Liu, W.J. Lu, L.J. Li, B.C. Zhao, and Y.P. Sun, Eur. Phys. J. B **85**, 414 (2012).

[84] R. Jha, V.P.S. Awana, Physica C **498**, 45 (2014).

[85] R. Jha, H. Kishan, V.P.S. Awana, J. Appl. Phys. **115**, 013902 (2014).

[86] R. Jha, V.P.S. Awana, Mat. Res. Exp. **1**, 016002 (2014).

[87] A. Kumar, R. Jha, S. K. Singh, J. Kumar, P. K. Ahluwalia, R. P. Tandon, and V. P. S. Awana, J. App. Phy. **111**, 033907 (2012).

[88] R. Jha, V.P.S. Awana, J Supercond Nov Magn. **27**, 1 (2014).

[89] S. Demura, K. Deguchi, Y. Mizuguchi, K. Sato, R. Honjyo, A. Yamashita, T. Yamaki, H. Hara, T. Watanabe, S. J. Denholme, M. Fujioka, H. Okazaki, T. Ozaki, O. Miura, T. Yamaguchi, H. Takeya, Y. Takano, arXiv:1311.4267.

[90] T. Sugimoto, B. Joseph, E. Paris, A. Iadecola, T. Mizokawa, S. Demura, Y. Mizuguchi, Y. Takano, N. L. Saini, Phys. Rev. B **89**, 201117(R) (2014).

[91] L. Li, Y. Li, Y. Jin, H. Huang, B. Chen, X. Xu, J. Dai, L. Zhang, X. Yang, H. Zhai, G. Cao and Z. Xu, Phys. Rev. B **91,** 014508 (2015).

[92] J. Liu, D. Fang, Z. Wang, J. Xing, Z. Du, X. Zhu, H. Yang and H. H. Wen, EPL, **106,** 67002 (2014).

[93] M. Nagao, S. Demura, K. Deguchi, A. Miura, S. Watauchi, T. Takei, Y. Takano, N. Kumada, I. Tanaka, J. Phys. Soc. Jpn. **82**, 113701 (2013).

[94] M. Nagao, A. Miura, S. Demura, K. Deguchi, S. Watauchi, T. Takei, Y. Takano, N. Kumada, I. Tanaka, Solid State Communications **178**, 33 (2014).

[95] L. Jiao, Z. F. Weng, J. Z. Liu, J. L. Zhang, G. M. Pang, C. Y. Guo, F. Gao, X. Y. Zhu, H. H. Wen, H. Q. Yuan, J. Phys.: Condens. Matter **27,** 225701 (2015).

[96] P. K. Biswas, A. Amato, C. Baines, R. Khasanov, H. Luetkens, Hechang Lei, C. Petrovic, E. Morenzoni Phys. Rev. B **88**, 224515 (2013).

[97] A. Gurevich, Rep. Prog. Phys. **74**, 124501 (2011).

[98] J. Jaroszynski, F. Hunte, L. Balicas, Jo Youn-jung, I. RaiCevic, A. Gurevich, D.C. Larbalestier, F.F. Balakirev, L. Fang, P. Cheng, Y. Jia, H.H. Wen, Phys. Rev. B **78**, 174523 (2008).

[99] P. J. Baker, S. R. Giblin, F. L. Pratt, R. H. Liu, G. Wu, X. H. Chen, M. J. Pitcher, D. R. Parker, S. J. Clarke and S. J. Blundell, New J. Phys. **11**, 025010 (2009).

[100] K. Deguchi, Y. Mizuguchi, S. Demura, H. Hara, T. Watanabe, S. J. Denholme, M. Fujioka, H. Okazaki, T. Ozaki, H. Takeya, T. Yamaguchi, O. Miura, Y. Takano, EPL **101**, 17004 (2013).

[101] Y. Mizuguchi, A. Miyake, K. Akiba, M. Tokunaga, J. Kajitani, O. Miura Phys. Rev. B **89**, 174515 (2014).

[102] N. R. Werthamer, E. Helfand, P. C. Hohenberg, Phys. Rev. **147**, 295 (1966).